\documentclass[prl,twocolumn,floatix,amssymb,showpacs,amsmath,superscriptaddress]{revtex4-1}
\pdfoutput=1
\usepackage{graphicx}
\usepackage{epsfig}
\usepackage{amsmath,amssymb}
\usepackage{ dsfont }
\usepackage{color}

\newcommand{\ket}[1]{\ensuremath{\left|{#1}\right\rangle}}
\newcommand{\bra}[1]{\ensuremath{\left\langle{#1}\right|}}

\begin{document}

\title{ Non-Abelian $SU(2)$ Lattice Gauge Theories in Superconducting Circuits}

\date{\today}

\author{A. Mezzacapo}
\affiliation{Department of Physical Chemistry, University of the Basque Country UPV/EHU, Apartado 644, 48080 Bilbao, Spain}

\affiliation{IBM T.J. Watson Research Center, Yorktown Heights, NY 10598, USA}

\author{E. Rico}
\affiliation{Department of Physical Chemistry, University of the Basque Country UPV/EHU, Apartado 644, 48080 Bilbao, Spain}
\affiliation{IKERBASQUE, Basque Foundation for Science, Maria Diaz de Haro 3, 48013 Bilbao, Spain}

\author{C. Sab\'in}
\affiliation{School of Mathematical Sciences, University of Nottingham, Nottingham NG7 2RD, United Kingdom}

\author{I. L. Egusquiza}
\affiliation{Department of Theoretical Physics and History of Science, University of the Basque Country UPV/EHU, Apartado 644, 48080 Bilbao, Spain}

\author{L. Lamata}
\affiliation{Department of Physical Chemistry, University of the Basque Country UPV/EHU, Apartado 644, 48080 Bilbao, Spain}

\author{E. Solano}
\affiliation{Department of Physical Chemistry, University of the Basque Country UPV/EHU, Apartado 644, 48080 Bilbao, Spain}
\affiliation{IKERBASQUE, Basque Foundation for Science, Maria Diaz de Haro 3, 48013 Bilbao, Spain}

\begin{abstract}
We propose a digital quantum simulator of non-Abelian pure-gauge models with a superconducting circuit setup. Within the framework of quantum link models, we build a minimal instance of a pure $SU(2)$ gauge theory, using triangular plaquettes involving geometric frustration. This realization is the least demanding, in terms of quantum simulation resources, of a non-Abelian gauge dynamics. We present two superconducting architectures that can host the quantum simulation, estimating the requirements needed to run possible experiments. The proposal establishes a path to the experimental simulation of non-Abelian physics with solid-state quantum platforms.
\end{abstract}


\maketitle

Gauge invariance is a central concept in modern physics, being at the core of the standard model of elementary particle physics. In particular, invariances with respect to $SU(2)$ and $SU(3)$ gauge symmetries characterize the weak interaction and quantum chromodynamics~\cite{munster1994, rothe2012}. In this sense, gauge theories represent a cornerstone in our understanding of the physical world and lie at the heart of diverse phenomena, such as the quark-gluon plasma or quantum spin liquids. In condensed matter physics, $SU(2)$ gauge fields can also emerge dynamically in relation to exotic many-body phenomena, like quantum Hall systems, frustrated magnets, or superconductors~\cite{affleck88,dagotto88,wen2004}.
 
Lattice gauge theories (LGT) are non-perturbative discrete formulations that contribute to the analysis of key features of these models, such as color confinement or chiral symmetry breaking. Starting from the seminal work by Wilson in 1974 \cite{Wilson1974, Kogut75}, LGT have attracted a significant attention across several branches of theoretical physics. In the last decades~\cite{durr2008}, quantum Montecarlo simulations have achieved unprecedented accuracies in determining the whole hadronic spectrum of the standard model. However, understanding its full phase diagram from first principles, or simulating dynamical processes, remain out of reach of current numerical computations. 

Quantum simulators~\cite{Feynman82} provide a new approach to solve complex long-standing problems in quantum physics. In a quantum simulator, a proper encoding of LGT can allow for the retrieval of information about ground state and dynamics, in a wide range of regimes. Previous works have considered Abelian~\cite{Weimer2010,Kapit2011,Zohar2011,Weimer2012,Banerjee2012,Zohar2013,Tagliacozzo2013,Zohar2013a,Marcos2014,Stannigel2014,Glaetzle2014,Notarnicola2015,Bazavov2015} and non-Abelian LGT in optical lattices~\cite{Zohar13,Tagliacozzo12, Banjeree13} (see also \cite{wiese14,zohar15} and references therein), both as analog and digital simulations~\cite{Georgescu14}. In these implementations, matter-gauge interactions are modeled as a second-order process from Hubbard-like interactions, while the simulation of pure-gauge dynamics remains more demanding.

In the last years, superconducting circuits have proven to be reliable devices that can host quantum information and simulation processes~\cite{Devoret13}. The possibility to perform quantum gates with high fidelities, together with high coherence times, makes them ideal devices for the realization of digital quantum simulations~\cite{LasHeras14,Salathe15,LasHeras15,Barends15,BarendsAdiabatic}, previously considered in ion-trap systems~\cite{Lanyon11,Casanova12,Mezzacapo12}.

\begin{figure}
\includegraphics[scale=0.283]{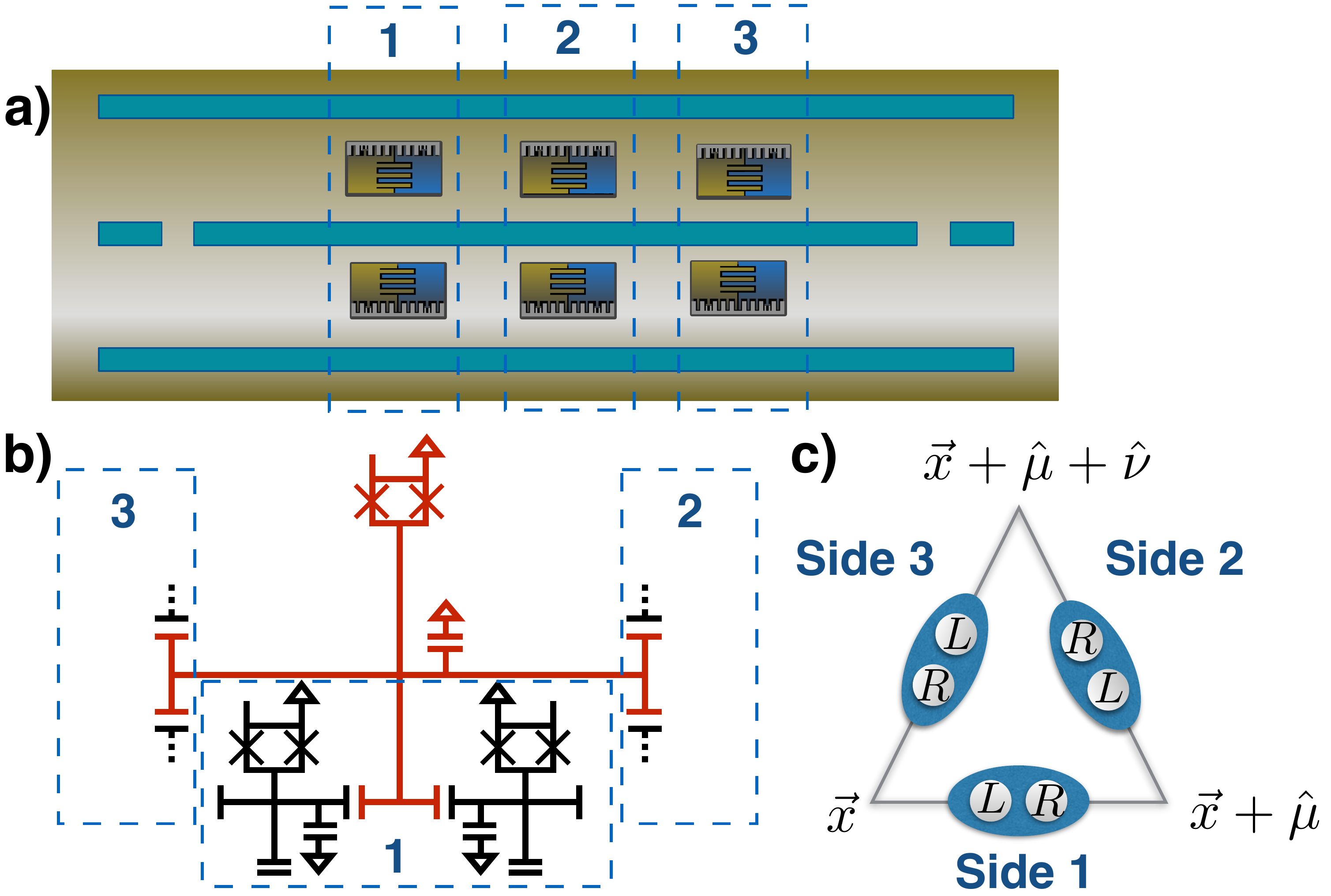}
\caption{(Color online) a) Six tunable-coupling transmon qubits coupled to a single microwave resonator. b) Six Xmon qubits on a triangular geometry, coupled to a central one. The box 1 in the scheme is implicitly repeated for the sides~2~and~3. Both setups can encode the dynamics of the $SU(2)$ triangular plaquette model schematized in c), where the left and right gauge degrees of freedom are explicitly depicted.\label{scheme}}
\end{figure}

In this Letter, we propose a digital quantum simulation of a non-Abelian dynamical $SU(2)$ gauge theory in a superconducting device. We start by building a minimal setup, based on a triangular lattice, that can encode pure-gauge dynamics. The degrees of freedom of a single triangular plaquette of this lattice are encoded into qubits. We propose two implementations of this quantum simulator, using two different superconducting circuit architectures, as depicted in Fig.~\ref{scheme}. We consider a setup in which six tunable-coupling transmon qubits are coupled to a single microwave resonator, and a device where six capacitively-coupled Xmon qubits stand on a triangular geometry, coupled to a central auxiliary one. We compute experimental requirements necessary to perform the simulation on one plaquette, and provide arguments for scaling to large lattices.    

{\it Lattice gauge theories.- }
LGT are discretized versions of a gauge theory. In a conventional approach of lattice gauge theories, space-time is discretized while ensuring a covariant formulation of the theory. In quantum simulations, there is no direct access to the time direction, that is fixed and continuous. Hence, the equivalent Hamiltonian formulation of lattice models is used. In this case, the space is discretized and the action of a local gauge invariant Hamiltonian characterizes a continuous dynamical evolution. This fictitious asymmetry of the time direction forces us to define a physical Hilbert space or Gauss law of the model. We make use of a set of discretized space points ${\vec x}$, and $SU(N)$ operators $u({{\vec x},{\hat \mu}})$ associated to each link, connecting two adjacent sites $({\vec x},{\vec x}+{\hat \mu})$. A single $u({{\vec x},{\hat \mu}})$ operator has two implicit color indices $(\alpha, \beta)$,  $u_{\alpha \beta}$, that connect the $\alpha$-th component of the fermionic field at position ${\vec x}$ with the $\beta$-th component of the field at position ${\vec x}+{\hat \mu}$.

In order to construct a non-Abelian $SU(N)$ theory built out of these $u({{\vec x},{\hat \mu}})$ operators, one has to build a Hamiltonian that is invariant under a generic gauge transformation
\begin{equation}
u({{\vec x},{\hat \mu}})\rightarrow e^{i\theta^{a}({\vec x})\tau^{a}}u({\vec x}, {\hat {\mu}})e^{-i\theta^{a}({\vec x}+{\hat \mu})\tau^{a}},\label{Gauge}
\end{equation}
where $\tau^{a}$ are the $(N^{2}-1)$ generators of the $SU(N)$ algebra in the corresponding representation, and $\theta^a$ the relative phase angles. It is straightforward to verify that the following Hamiltonian on a square lattice is gauge invariant
\begin{equation}
H=J\sum_{{\vec x}} \text{Tr}[u({\vec x}, {\hat {\mu}}) u({\vec x}+{\hat \mu},{\hat \nu})u^{\dagger}({\vec x}+{\hat \nu},{\hat \mu}) u^{\dagger}({\vec x},{\hat \nu})] + \text{H.c.},
\label{WilsonLoop}
\end{equation}
where the trace is performed over the color indices, $J$ has energy units, and the directions $\hat{\mu}$ and $\hat{\nu}$ span the two-dimensional lattice. This Hamiltonian is a pure gauge operator, as it does not involve any fermionic operator and it is the minimum instance of a {\it Wilson loop} around a plaquette~\cite{munster1994, rothe2012}.

We focus now on the construction of a non-Abelian $SU(2)$ quantum link model that mimics this gauge-invariant behavior in a finite-dimensional Hilbert space, suitable for quantum simulations. Notice that the framework of the quantum link models is valid for any compact Lie group~\cite{Brouwer99}. We use a four-dimensional Hilbert space or, equivalently, two qubits associated with each link. One can define link operators $U({\vec x},{\hat \mu})$ acting on this four-dimensional Hilbert space with the associated gauge generators $G^{a}({\vec x})$ that satisfy the $SU(2)$ algebra 
\begin{equation}
[G^{a}({\vec x}),G^{b}({\vec y)}]=i\delta_{{\vec x}{\vec y}} \sum_{c} \epsilon_{abc}G^c(\vec{x}).
\end{equation}
To define the set of operators, we use a quantum link formulation~\cite{Brouwer99}, where the Hilbert space of a link is given by a set of bosonic modes $c_{\alpha j}({\vec x},{\hat \mu})$ that implement the Schwinger representation of the $SU(2)$ algebra, with $\alpha \in \{ \uparrow, \downarrow \}$, acting on two different sites $j \in \{L,R\}$ on the link between the two adjacent vertices ${\vec x}$ and ${\vec x}+{\hat \mu}$. In this space, we build two sets of {\it right} and {\it left} generators $R^{a}$, ${L}^{a}$, on this finite-dimensional Hilbert space, using the bosonic modes $c_{\alpha j}({\vec x},{\hat \mu})$~\cite{Supplementary}. We define the gauge generators as $G^{a}({\vec x})=\sum_{|{\hat \nu}|} L^{a}({\vec x},{\hat \nu})+ R^{a}({\vec x}-{\hat \nu},{\hat \nu})$, where the sum over $|\hat{\nu}|$ is taken among all the links of the lattice converging to a single vertex ${\vec x}$. 

\begin{figure}
\includegraphics[scale=0.3]{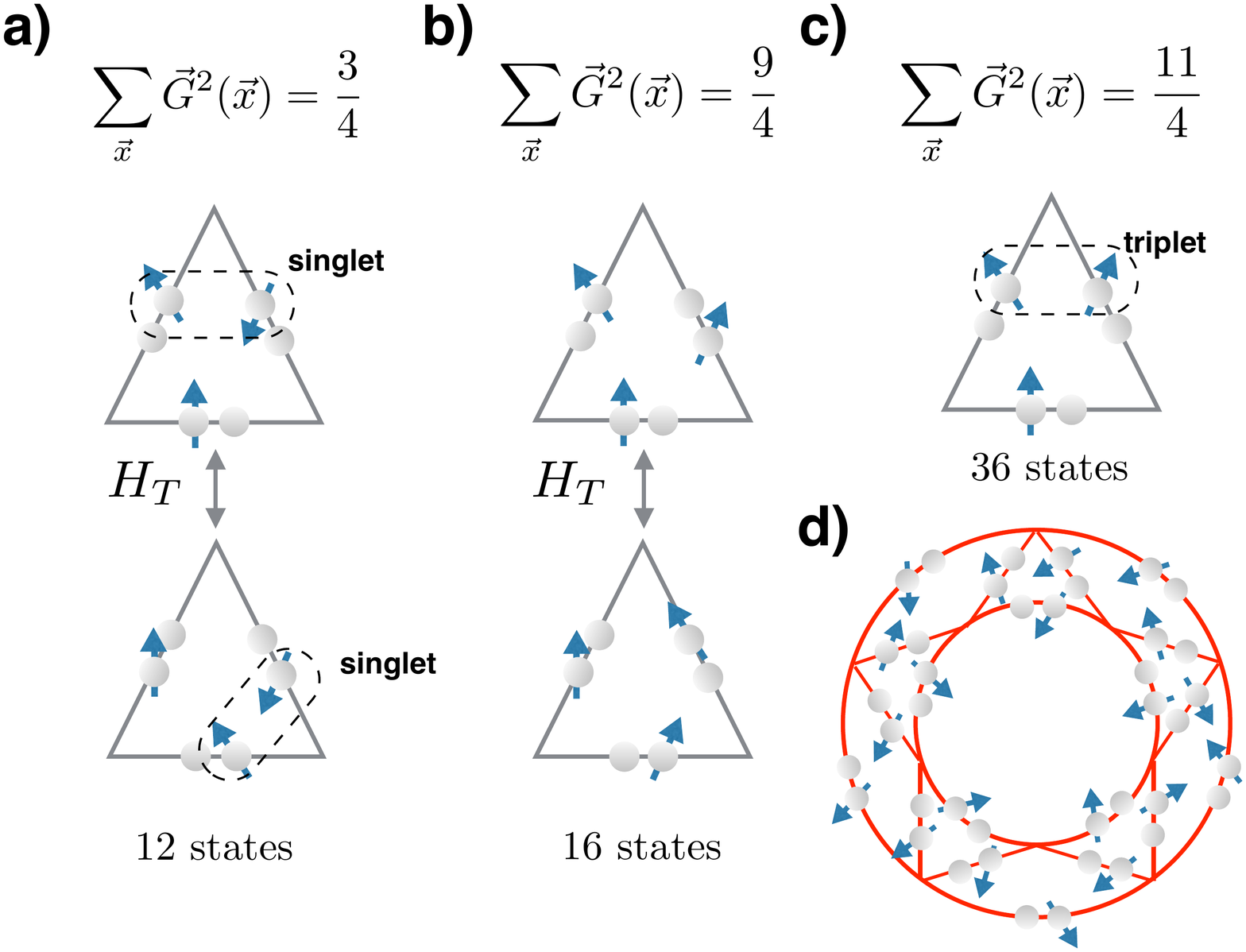}
\caption{(Color online) a), b), and c) Different gauge invariant sectors for a triangular plaquette, together with the sector degeneracy and the action of the Hamiltonian $H_T$ upon them. d) Extended lattice with periodic boundary conditions. This lattice allows for the $\sum_xG^2(x)=0$ sector.\label{States}}
\end{figure}

The representations of $SU(2)$ are quasi-real, therefore, ordinary and their dual representation (i.e., particle and antiparticle) are equivalent. Therefore, there are two multiplets with well-defined $SU(2)$ transformations~\cite{Supplementary}. One can finally define the link operators, acting on the finite-dimensional Hilbert space of one link, in terms of Schwinger bosons on a given link~\cite{Horn81,Orland90,mathur05},
\begin{equation}
\label{Udef}
U_{\alpha \beta}=
\begin{pmatrix}
c_{\uparrow L} & -i c^{\dagger}_{\downarrow L} \\
c_{\downarrow L} & i c^{\dagger}_{\uparrow L}
\end{pmatrix}
\cdot
\begin{pmatrix}
c^{\dagger}_{\uparrow R} & c^{\dagger}_{\downarrow R} \\
i c_{\downarrow R} & -i c_{\uparrow R} 
\end{pmatrix}.
\end{equation}
These operators $U$ have the following commutation rules with the left and right operators
\begin{equation}
\label{CommXiU}
\begin{split}
[U({\vec y},{\hat \nu}), R^{a}({\vec x},{\hat \mu})]&= - U({\vec y},{\hat \mu})\frac{\sigma^{a}}{2} \delta_{{\vec x}{\vec y}}\delta_{{\hat \nu}{\hat \mu}},\\ 
[U({\vec y},{\hat \nu}), L^{a}({\vec x},{\hat \mu})]&=\frac{\sigma^{a}}{2}U({\vec y},{\hat \mu})\delta_{{\vec x}{\vec y}}\delta_{{\hat \nu}{\hat \mu}},
\end{split}
\end{equation}
where we have defined the usual Pauli matrices $\sigma^a$, see also~\cite{Supplementary}, while $\sigma^0$ is defined as the identity operator. The commutation rules for a general gauge transformation follow in a straightforward way,
\begin{flalign}
\prod_{\vec y}e^{-i\theta^{a}({\vec y})G^{a}(\vec y)}&U({\vec x}, {\hat {\mu}})\prod_{\vec z}e^{i\theta^{a}({\vec z})G^{a}(\vec z)}\hspace{2cm}\nonumber \\
&=e^{i\theta^{a}({\vec x}) \frac{\sigma^{a}}{2}}U({\vec x}, {\hat {\mu}})e^{-i\theta^{a}({\vec x}+{\hat \mu})\frac{\sigma^{a}}{2}},
\label{CommUG}
\end{flalign}
where the products over $\vec{y}$ and $\vec{z}$ are extended over the whole lattice. This equation shows that quantum link models are formulations of gauge-invariant models. In fact, one can notice how the action of a gauge transformation in Eq.~(\ref{CommUG}) mimics Eq.~(\ref{Gauge}).

A generic state in the local Hilbert space $| n_{\uparrow L} , n_{\downarrow L} , n_{\uparrow R} , n_{\downarrow R} \rangle$ of a quantum link is given by the occupation of the operators $c^\dag_{\uparrow(\downarrow),R(L)}$. Since the total occupation per link, $n_{\uparrow L} + n_{\downarrow L} + n_{\uparrow R} + n_{\downarrow R}$, is a constant of motion [see Eq.~(\ref{Udef})], we restrict to the subspace with one occupied mode per link.

Notice that matter-gauge interactions in $1+1$ dimensions can be derived~\cite{Supplementary} in a second-order perturbation theory, considering extensions of previous works~\cite{hauke,rabl}. In the following, we will focus instead on pure-gauge two-dimensional interactions.

{\it Triangular lattice.-} The continuum limit of quantum link models and their thermodynamical properties are topics under active research. In related condensed matter models, the importance of the lattice geometry has been shown in fundamental aspect of the phase diagram, as the existence of confinement and de-confinement phases \cite{Rokhsar1988,Moessner2001}. Quantum simulations on large lattices may provide insights on the continuum limit of LGT, and its dependence on the geometry of the underlying lattice. In this spirit, we consider a minimal implementation of a pure $SU(2)$ invariant model in a triangular lattice, by using triangular plaquettes, as depicted in Fig.~\ref{scheme}c. In this case, the pure-gauge Hamiltonian on a single plaquette reads 
\begin{equation}
H_T=-J~\text{Tr}\left[U(\vec{x},\hat{\mu})U(\vec{x}+\hat{\mu},\hat{\nu})U(\vec{x}+\hat{\mu}+\hat{\nu},-\hat{\mu}-\hat{\nu})\right].
\end{equation}
This interaction corresponds to the magnetic term of a gauge invariant dynamics, which acts on closed loops. We focus on the magnetic term, since the representation of the electric counterpart is trivial.

Since this Hamiltonian commutes with gauge generators, $ [H_T, G^{a}(\vec{x}) ]=0, \,  ~ \forall \vec{x},a$, these are constant of motion, and one can define different gauge sectors with the initial values of the generators. Due to the triangular geometry and having one occupied mode per link, there are no global states that have zero eigenvalue for the gauge generators $G^{a}(\vec{x})$ at every vertex in a single plaquette. In general, the absence of the zero-eigenvalue sector depends on the topology of the lattice, and is avoided, for example, in the periodic lattice of Fig.~\ref{States}d. The different gauge sectors can be characterized by the eigenvalues of the operator $\sum_{\vec{x}} {\vec G^{2}({\vec x })}=\sum_{\vec{x}}\sum_a [G^a({\vec x })]^2$, as in Fig.~\ref{States}a, ~\ref{States}b, and~\ref{States}c, see the supplemental material for additional details~\cite{Supplementary}. 

The local Hilbert space of a link is four dimensional, and it can be faithfully spanned by two qubits, called ``position'' $\sigma^{a}_{\textrm{pos}}$ and ``spin'' qubit $\sigma^{a}_{m}$. In this subspace, it is useful to define the operators $\Gamma^{0} = \sigma^{x}_{\textrm{pos}} \sigma^{0}_{m}$, $\Gamma^{a} =\sigma^{y}_{\textrm{pos}} \sigma^{a}_{m}$, such that the total Hamiltonian is written as~ \cite{Supplementary}
\begin{eqnarray}
\label{HG}
H_{T}=&&-J\Big\{ \Gamma^0_{12}\Gamma^0_{23}\Gamma^0_{31}+ \sum_{abc}\epsilon_{abc}\Gamma^a_{12}\Gamma^b_{23}\Gamma^c_{31} \\
&&-\sum_{a}\Big[\Gamma^0_{12}\Gamma^a_{23}\Gamma^a_{31} + \Gamma^a_{12}\Gamma^0_{23}\Gamma^a_{31}+\Gamma^a_{12}\Gamma^a_{23}\Gamma^0_{31}\Big] \Big\} \nonumber.
\end{eqnarray}

\begin{figure}
\includegraphics[scale=1]{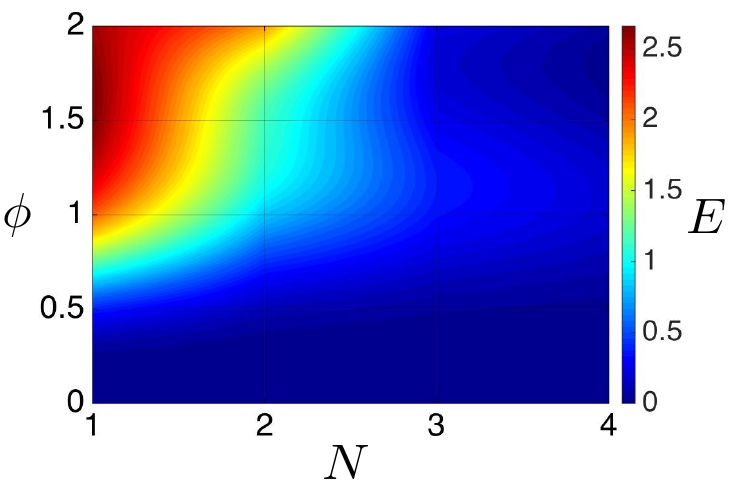}
\caption{(Color online) The relative gauge deviation of the digitally simulated gauge value ($\langle\rangle_D$), versus the ideal value ($\langle\rangle_I$), $E=\left(\langle\sum_{\vec{x}} {\vec G^{2}({\vec x })}\rangle_I-\langle\sum_{\vec{x}} {\vec G^{2}({\vec x })}\rangle_D\right)/ \langle\sum_{\vec{x}} {\vec G^{2}({\vec x })}\rangle_I$, as a function of the digital steps $N$ and the simulated phase $\phi=Jt$. The error decreases with large $N$, and small $\phi$. The contour plot is interpolated from numerical data at $N=\{1,2,3,4\}$. The initial state is chosen as depicted in Fig.~\ref{States}a.\label{ContourGauge}} 
\end{figure}

{\it Superconducting circuit implementation.-} In order to simulate the interaction of Eq.~(\ref{HG}), one can decompose its dynamics in terms of many-body monomials, and implement them sequentially with a digitized approximation~\cite{Lloyd96}. In a digital approach, one decomposes the dynamics of a Hamiltonian $H=\sum_{k=1}^mh_k$ by implementing its components stepwise, $e^{-iHt}\approx\left(\prod_{k=1}^m e^{-ih_kt/N}\right)^N$ (here and in the following $\hbar=1$), for a total of $m~\times~N$~gates, with an approximation error that goes to zero as the number of repetitions $N$ grows. In a practical experiment, each quantum gate $e^{-ih_kt}$ will be affected by a given error $\epsilon_k$. By piling up sequences of such gates, for small gate errors $\epsilon_k \ll 1$, the total protocol will be affected by a global error, which is approximately the sum $\epsilon\approx\sum_k\epsilon_k$. This model for error accumulation has been proved in recent experiments~\cite{Salathe15,Barends15}. 

The effects of digitization in the simulation of LGT can be observed in Fig.~\ref{ContourGauge}. We have numerically integrated a Schr\"odinger equation regulated by the Hamiltonian in Eq.~(\ref{HG}), choosing as initial state one of the $12$ states shown in Fig.~\ref{States}a, for different simulated phases $\phi=Jt$~\cite{Supplementary}. We compute both the evolution with the ideal Hamiltonian and with a digital sequence, in which we act with a single monomial at a time, repeating the protocol $N$ times. As a figure of merit, we compute the relative deviation $E=\left(\langle\sum_{\vec{x}} {\vec G^{2}({\vec x })}\rangle_I-\langle\sum_{\vec{x}} {\vec G^{2}({\vec x })}\rangle_D\right)/ \langle\sum_{\vec{x}} {\vec G^{2}({\vec x })}\rangle_I$ from the ideal value of the gauge invariant, where $\langle\rangle_{I(D)}$ stand for average values computed on the ideal (digitally simulated) state. The deviation goes to zero for large $N$ and small phases, defining surfaces at given error tolerance in the $N$-$\phi$ space. 

To simulate the pure-gauge interaction in a single triangular plaquette, we first consider a setup in which six tunable-coupling transmon qubits are coupled to a single microwave resonator~\cite{Gambetta11,Srinivasan12}. Each tunable-coupling qubit is built using three superconducting islands, connected by two SQUID loops. Acting on these loops with magnetic  fluxes, one can modify the coupling of the qubits with the resonator, without changing their transition frequencies. For further clarifications of the experimental setup involving tunable-coupling transmon qubits, see the Supplemental Material~\cite{Supplementary}. By threading with magnetic fluxes at high frequencies, one can drive simultaneous red and blue detuned sidebands, and perform collective gates~\cite{Mezzacapo14}. Each many-body operator can be realized as a sequence of collective and single-qubit gates, 
\begin{equation}
\label{UN}
U_N=e^{i(\pi/4)\sum_{i<j}\sigma^x_i\sigma^x_j}e^{i\phi\sigma_1^z}e^{-i(\pi/4)\sum_{i<j}\sigma^x_i\sigma^x_j},
\end{equation}
where the indices $i$ and $j$ run from $1$ to $N$. In this way, one can implement a generic (up to $2N$ local rotations) $N$-body evolution operator $U_N\equiv e^{-i\phi\sigma^m_1\sigma^n_2\cdots\sigma^k_N}$, with $\{m,n,...k\}\in\{x,y,z\}$, using $2$ collective gates, and a number of single qubit gates which is upper bounded by $2N+1$, counting the single qubit gate in Eq.~(\ref{UN}) and the $2N$ rotations necessary to map $U_N$ to any $N$-body operator. The simulation for one digital step will amount to implementing $32$ collective gates and a number of single qubit gates which is upper bounded by $184$~\cite{Supplementary}.

\begin{figure}
\includegraphics[scale=0.6]{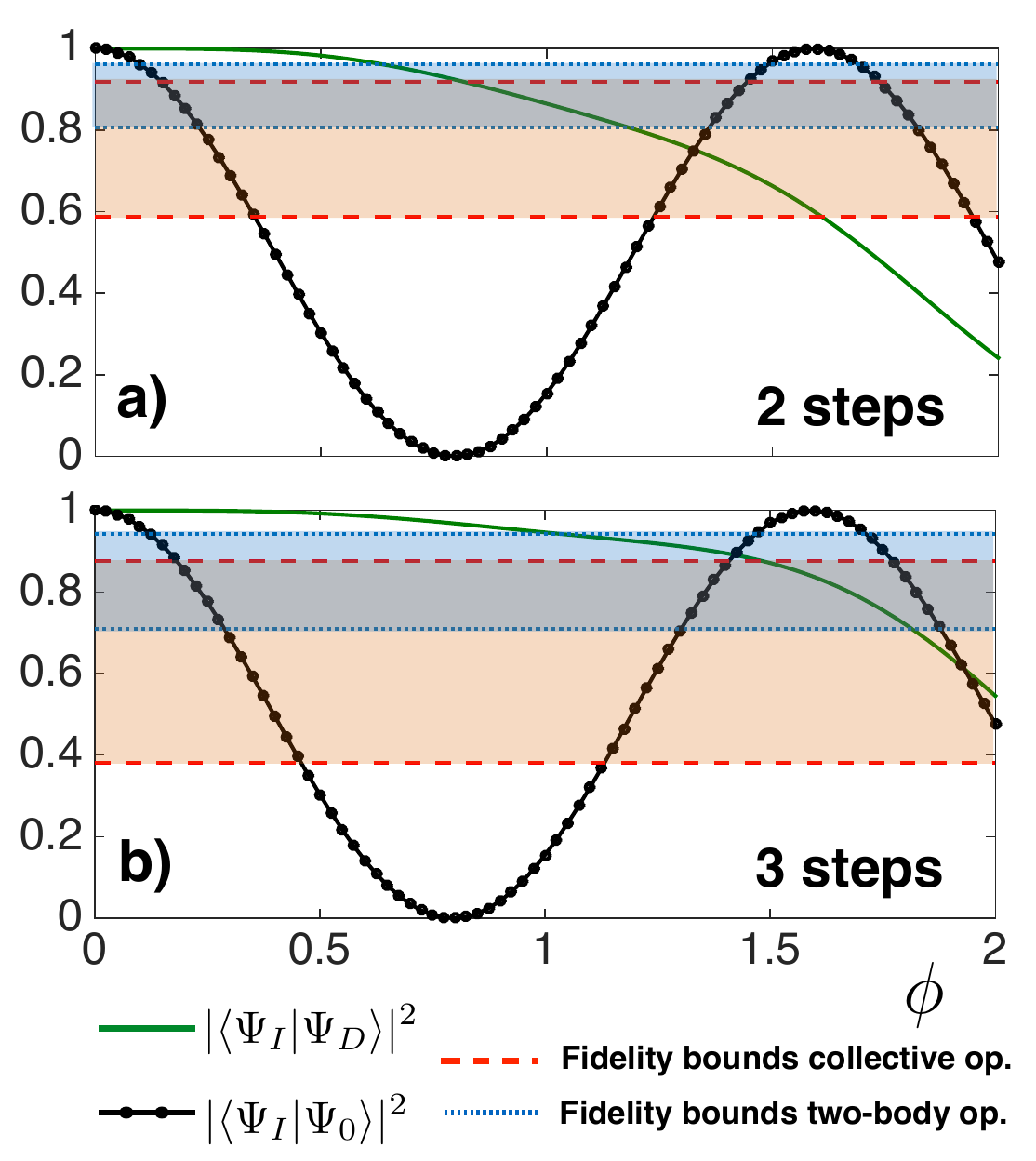}
\caption{(Color online) Digital quantum simulation of pure gauge dynamics for a) $2$ and b) $3$ digital steps. The initial state $\ket{\Psi_0}$ is chosen as in Fig.~\ref{States}b. Periodic oscillations between the position and qubit degrees of freedom at the vertices are witnessed by $|\langle\Psi_I|\Psi_0\rangle|^2$. The theoretical fidelity $|\langle\Psi_I|\Psi_D\rangle|^2$ increases with the number of digital steps, together with the width of the fidelity bands due to experimental imperfections. \label{Simulationerr}}
\end{figure}

We consider now the architecture of Fig.~\ref{scheme}b, in which six Xmon qubits in a triangular geometry are capacitively coupled with an additional central ancillary qubit~\cite{Barends13}. In this case, the collective interactions $U_{N}$ can be decomposed and performed with pairwise $C$-phase gates, using the central ancillary qubit to mediate non-nearest interactions. In this way, the quantum simulation of one digital step of the Hamiltonian in Eq.~(\ref{HG}) will amount to realize \mbox{$168$~$C$-Phase~gates} and a number of single-qubit rotations which is upper bounded by $520$. For additional details on the experimental setup and gate counts, see~\cite{Supplementary}. 

To estimate the effects of experimental imprefections on the simulations, we plot in Fig.~\ref{Simulationerr} the dynamics of a single plaquette, starting from a \mbox{$\sum_x\vec{G}_x^2=9/4$} state. The observed periodic dynamical oscillations between the two paired states, depicted in Fig.~\ref{States}b, are signaled by the periodic behavior of the overlap $|\langle\Psi_I|\Psi_0\rangle|^2$. The simulation is run for $N=2$ and $N=3$ steps, necessary to observe one flip between the two states and one full oscillation, respectively. On top of these oscillations, we plot fidelity bands that estimate the fidelity cap due to accumulated gate errors in the digital protocol, in the case of the two presented setups in Fig.~\ref{scheme}, with collective and two-qubit gates. In order to see coherent oscillations, the fidelity of the two-body gates has to be one order of magnitude better with respect to the collective ones, with an error window of approximately $0.05-0.01\%$ in the collective gate case and a $0.005-0.001\%$ for the $C$-phase gate setup. We assume a factor $1/20$ for the error loss on a single qubit gate. From the plots, it can be seen that, while the digital fidelity $|\langle\Psi_I|\Psi_D\rangle|^2$ increases, the fidelity bands due to experimental errors broaden, affecting the simulation. The intersection of the digital fidelity line with the bands marks a regime dominated by experimental imperfections, for small $\phi$, and one in which the error of the digital expansion prevails, for large $\phi$.           

The proposed simulation can be extended to larger lattices, where one could analyze dynamical properties of the confinement-deconfinement phase transition, which can be traced back to the breaking of the global center symmetry~\cite{greensite} in groups with a non-trivial center group, like the $SU(2)$ case. Moreover, such a LGT simulator can be used to perform quench experiments in a ladder configuration, analyzing breaking of gluon strings between pairs of particles and anti-particles~\cite{pepe,kuhn,pichler}.

In order to scale the quantum simulation to large qubit lattices, one has to consider that the accumulated gate error does not depend on the size of the lattice~\cite{Salathe15,Barends15}, and only on the number of gates. The quantum resources necessary to run the simulation scale polynomially in the lattice size, and sub-polynomially in the digital error~\cite{Berry}, making the whole protocol efficient. With a current total number of gates that exceeds 1000~\cite{Barends15,BarendsAdiabatic}, we believe that in the near future simulations of LGT on large scales will be feasible. Furthermore, the possibility of performing quantum error correction for digital quantum simulations may drastically increase its effectiveness~\cite{BarendsN}.  

AM and LL acknowledge useful discussions with Rami Barends. ER thanks D. Banerjee, T. Calarco, M. Dalmonte, S. Montangero, U.-J. Wiese, P. Zoller for discussions, in the framework of quantum technologies for lattice gauge theories. We acknowledge financial support from Basque Government Grants IT472-10 and IT559-10, Spanish MINECO FIS2012-36673-C03-02, Ram\'on y Cajal Grant RYC-2012-11391, UPV/EHU Project No. EHUA14/04, UPV/EHU UFI 11/55, PROMISCE and SCALEQIT European projects.

\pagebreak

\begin{widetext}

\section{SUPPLEMENTAL MATERIAL}

In this supplemental material, we perform additional analysis that support the results obtained in the main article. 

\subsection{Explicit $SU(2)$ gauge operators in terms of Schwinger modes}

In this section, we explicitly derive the link operators in terms of Schwinger bosons. Following~\cite{Brouwer99S}, we define the operators $R^{a}$, ${L}^{a}$ as {\it right} and {\it left} generators, acting on the finite-dimensional Hilbert space of the link,
\begin{equation}
\begin{split}
R^{a}({\vec x},{\hat \mu})=& \sum_{\{\alpha ,\beta \} \in \{ \uparrow, \downarrow \} }c^{\dagger}_{\alpha R}({\vec x},{\hat \mu})  \frac{\sigma^{a}_{\alpha \beta}}{2} c_{\beta R}({\vec x},{\hat \mu}),\\
L^{a}({\vec x},{\hat \mu})=& \sum_{\{\alpha ,\beta \} \in \{ \uparrow, \downarrow \} }c^{\dagger}_{\alpha L}({\vec x},{\hat \mu})  \frac{\sigma^{a}_{\alpha \beta}}{2} c_{\beta L}({\vec x},{\hat \mu}),
\end{split}
\end{equation}
where $c_{\alpha j}$ are bosons that implement the Schwinger representation of the $SU(2)$ algebra, and the Pauli matrices 
\begin{equation} 
\sigma^{(0)}= \begin{pmatrix} 1 & 0 \\ 0 & 1 \end{pmatrix}; ~ \sigma^{(1)}= \begin{pmatrix} 0 & 1 \\ 1 & 0 \end{pmatrix}; ~ \sigma^{(2)}= \begin{pmatrix} 0 & -i \\ i & 0 \end{pmatrix}; ~ \sigma^{(3)}= \begin{pmatrix} 1 & 0 \\ 0 & -1 \end{pmatrix}.
\end{equation}
The bosonic operators act on two different sites $p \in \{L,R\}$ on the link between the two adjacent sites ${\vec x}$ and ${\vec x}+{\hat \mu}$ and obey the $SU(2)$ commutation rules
\begin{equation}
\begin{split}
[R^{a},R^{b}]=i \sum_{c} \epsilon^{abc}R^{c}, ~ \, ~[R^{a},L^{b}]=0, ~ \, ~ [L^{a},L^{b}]=&i \sum_{c} \epsilon^{abc}L^{c}, \\
\end{split}
\end{equation}
taking into account that they commute on different links (i.e., different ${\vec x}$, ${\hat \mu}$). With the Schwinger representation, there are two multiplets with well-defined $SU(2)$ commutation relations
\begin{equation}
\begin{split}
\left[c_{\alpha L} , L^{a}\right] =\sum_{\gamma} \frac{\sigma^{a}_{\alpha \gamma}}{2} c_{\gamma L}, ~ \, ~ &\left[ \left( \sum_{\beta} \sigma^{y}_{\alpha \beta}c^{\dagger}_{\beta L} \right), L^{a}\right] =\sum_{\gamma} \frac{\sigma^{a}_{\alpha \gamma}}{2} \left(\sum_{\beta} \sigma^{y}_{\gamma \beta} c^{\dagger}_{\beta L} \right),\\
\left[c^{\dagger}_{\alpha R} , R^{a}\right] = - \sum_{\gamma} c^{\dagger}_{\gamma R} \frac{\sigma^{a}_{\gamma \alpha}}{2}, ~ \, ~ &\left[\left( \sum_{\beta} c_{\beta R}   \sigma^{y}_{\beta \alpha} \right), R^{a}\right] = -\sum_{\gamma} \left(\sum_{\beta} c_{\beta L} \sigma^{y}_{ \beta \gamma} \right)  \frac{\sigma^{a}_{ \gamma \alpha}}{2}.
\end{split}
\end{equation}
Given these two multiplets, we can define the link operator as
\begin{equation}
U_{\alpha \beta} =  c_{\alpha L} c^{\dagger}_{\beta R} + \sum_{\mu \nu} \sigma^{y}_{\alpha \mu} c^{\dagger}_{\mu L} c_{\nu R} \sigma^{y}_{\nu \beta}.
\end{equation}

We stress that these quantum link operators, or gauge fields, are associated with links between lattice points, transporting color information between the two points they connect. The two indices of the link operator $U_{\alpha \beta}({\vec x},{\hat \mu})$ are identified with the two ends of the link $({\vec x},{\hat \mu})$. The left (right) index is associated with the beginning (end) of the link as shown in Fig. \ref{gaugecartoon}. The gauge transformation acts on $U({\vec x},{\hat \mu})$ according to
\begin{equation}
U({\vec x}, {\hat {\mu}}) =e^{i \sum_{a} \theta^{a}({\vec x}) \frac{\sigma^{a}}{2}}U({\vec x}, {\hat {\mu}})e^{-i \sum_{a} \theta^{a}({\vec x}+{\hat \mu})\frac{\sigma^{a}}{2}}.
\end{equation}
Since in general $\theta^{a}({\vec x})$ and $\theta^{a}({\vec x}+{\hat \mu})$ are different, our gauge-invariant equations will require invariance under right multiplication and left multiplication separately. An arbitrary gauge transformation can be built from individual gauge transformations at the point of the lattice. Therefore, it is sufficient to consider just a gauge transformation at the lattice site $\vec{x}$. Every link emanating from the point is affected by the gauge transformation generated by
\begin{equation}
G^{a} \left( \vec{x} \right) = \sum_{{\hat {\mu}}} L^{a}\left(\vec{x}-{\hat {\mu}}, {\hat {\mu}}\right) + R^{a} \left( \vec{x}, {\hat {\mu}} \right),
\end{equation}
and the total color $Q^{a} \left( \vec{x}, {\hat {\mu}} \right)$ carried by a link is just the difference: $R^{a} \left( \vec{x}, {\hat {\mu}} \right)- L^{a} \left( \vec{x}, {\hat {\mu}} \right)$.

The sum over non-Abelian electric fields emanating from a single point of the lattice is the lattice analog of the electromagnetic divergence of the electric field,  $\vec{\nabla} \cdot \vec{E}^{a} $. However, because the electric field itself varies from $L^{a} \left( \vec{x}, {\hat {\mu}}\right)$ to $R^{a} \left( \vec{x}, {\hat {\mu}}\right)$ along the link, there is an additional contribution given by the total color carried by the link. Thus, the generator is recast into: $G^{a} \left( \vec{x} \right) = \vec{\nabla} \cdot \vec{E}^{a} \left( \vec{x} \right) - \frac{1}{2} \sum_{{\hat {\mu}}} Q^{a} \left( \vec{x}, {\hat {\mu}} \right)$. The gauge invariance of the physical sector is defined by  $G^{a} \left( \vec{x} \right) |\Psi \rangle = 0$, i.e., at every vertex the state transforms locally as a singlet of $SU(2)$. If we identify the local color density $\rho^{a}\left( \vec{x}, {\hat {\mu}} \right) \equiv \frac{1}{2} \sum_{{\hat {\mu}}} Q^{a} \left( \vec{x}, {\hat {\mu}} \right)$, then $\vec{\nabla} \cdot \vec{E}^{a} \left( \vec{x} \right) = \rho^{a}\left( \vec{x}, {\hat {\mu}} \right)$ is the usual Gauss' law. These are properties of non-Abelian lattice gauge theories, independent of their representation via quantum link models.

\subsection{Local Hilbert space of $SU(2)$ quantum links}

\begin{figure}
\includegraphics[scale=0.46]{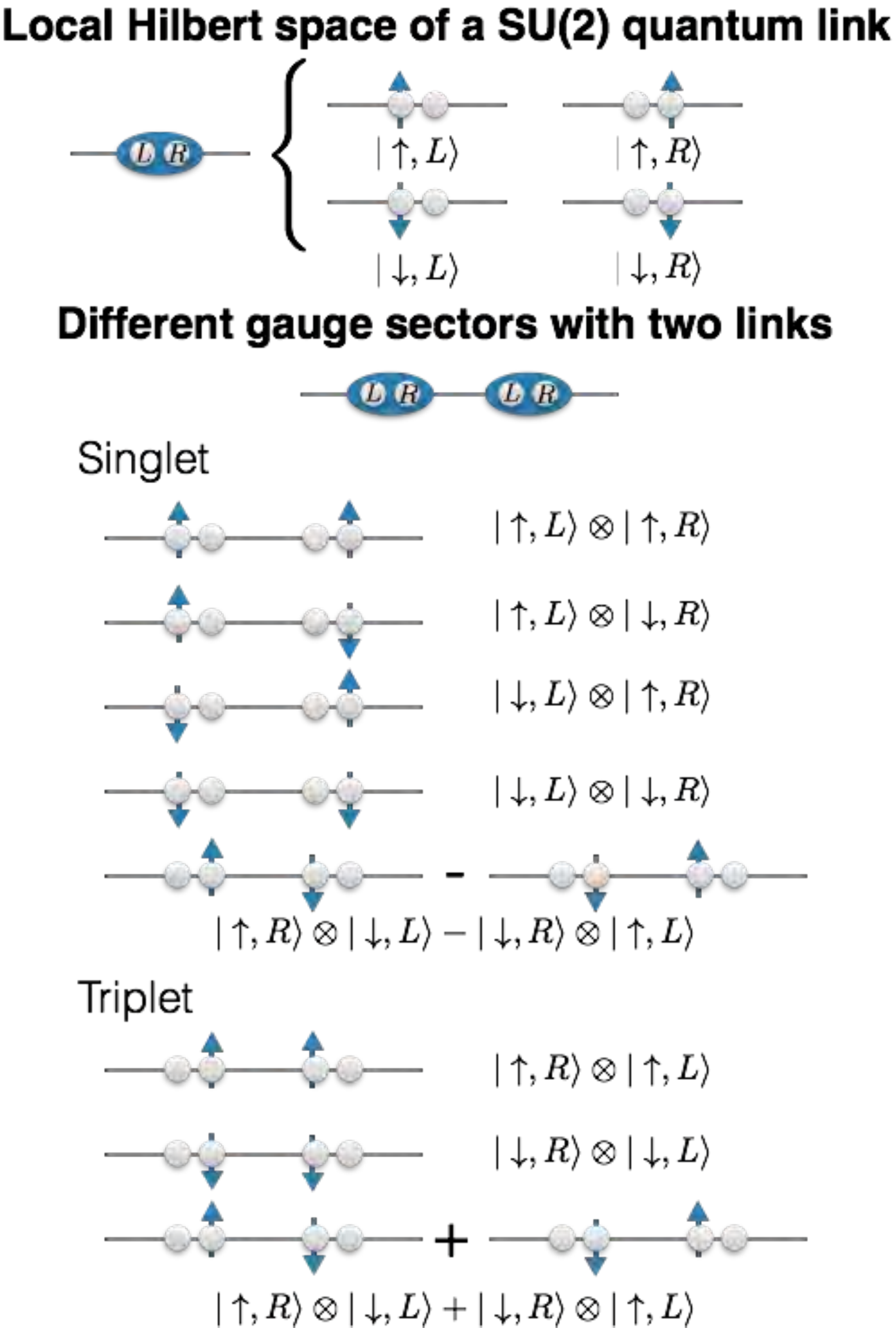}
\caption{(Color online) The four dimensional Hilbert space of the minimal representation of a $SU(2)$ quantum link. Below, the two different gauge sectors in the Hilbert space of two quantum links, i.e. the singlet and the triplet sector. \label{gaugecartoon}}
\end{figure}

The local degrees of freedom in a $SU(2)$ link are encoded in the quantum state of four different bosonic modes $| n_{\uparrow L} , n_{\downarrow L} , n_{\uparrow R} , n_{\downarrow R} \rangle$. Different representations of this local Hilbert space are given by the occupation of these modes. The Hilbert space with one excitation in total is four-dimensional, and it can be decomposed in a spin degree of freedom with the $SU(2)$ symmetry group and a position degree of freedom.

A $SU(2)$ gauge invariant interaction commutes with the local generator of the $SU(2)$ at any vertex. As a consequence, the total angular momentum at any vertex of the lattice is a conserved quantity of the dynamics. In the simplest example of two quantum links, i.e. just one vertex, there are only two possible gauge sectors, the singlet sector where the total angular momentum around the vertex is zero, and the triplet sector where the total angular momentum is equal to one.

\subsection{Reduction to qubits - Triangular lattice}

In the main text, we have considered the local four-dimensional Hilbert space spanned by the occupation of four different modes $| n_{\uparrow L} , n_{\downarrow L} , n_{\uparrow R} , n_{\downarrow R} \rangle$. By choosing the subspace with one excitation per link, we can faithfully represent its quantum state with two qubits, the position and spin qubits, $|\textrm{pos},m^{z} \rangle$. The excitations of these qubits are related to the left and right modes through $\textrm{pos} = \left( n_{\uparrow R} + n_{\downarrow R} \right) - \left(n_{\uparrow L} + n_{\downarrow L} \right)$ and $m^{z}= \left(n_{\uparrow R} +  n_{\uparrow L} \right)- \left( n_{\downarrow R} + n_{\downarrow L} \right)$.

With these definitions, the left and right operators defined in the main text are recast into
\begin{equation}
\begin{split}
R^{a}=& \sum_{\{\alpha ,\beta \} \in \{ \uparrow, \downarrow \} }c^{\dagger}_{\alpha R}  \frac{\sigma^{a}_{\alpha \beta}}{2} c_{\beta R} = \frac{\sigma^{0}_{\textrm{pos}} + \sigma^{z}_{\textrm{pos}}}{2} \frac{\sigma^{a}_{m}}{2},\\
L^{a}=&  \sum_{\{\alpha ,\beta \} \in \{ \uparrow, \downarrow \} }c^{\dagger}_{\alpha L}  \frac{\sigma^{a}_{\alpha \beta}}{2} c_{\beta L} =\frac{\sigma^{0}_{\textrm{pos}} - \sigma^{z}_{\textrm{pos}}}{2} \frac{\sigma^{a}_{m}}{2},\\
U=&\begin{pmatrix}
c_{\uparrow L} c^{\dagger}_{\uparrow R} + c^{\dagger}_{\downarrow L}c_{\downarrow R} & c_{\uparrow L} c^{\dagger}_{\downarrow R} - c^{\dagger}_{\downarrow L} c_{\uparrow R}\\
c_{\downarrow L} c^{\dagger}_{\uparrow R} - c^{\dagger}_{\uparrow L} c_{\downarrow R} & c_{\downarrow L} c^{\dagger}_{\downarrow R} + c^{\dagger}_{\uparrow L} c_{\uparrow R}
\end{pmatrix} = 
\frac{1}{2} \begin{pmatrix}
\sigma^{x}_{\textrm{pos}} \sigma^{0}_{m} + i \sigma^{y}_{\textrm{pos}} \sigma^{z}_{m} & i \sigma^{y}_{\textrm{pos}} \left( \sigma^{x}_{m} - i \sigma^{y}_{m} \right) \\
i \sigma^{y}_{\textrm{pos}} \left( \sigma^{x}_{m} + i \sigma^{y}_{m} \right) & \sigma^{x}_{\textrm{pos}} \sigma^{0}_{m} - i \sigma^{y}_{\textrm{pos}} \sigma^{z}_{m}
\end{pmatrix}\\
= &\frac{1}{2} \left[ \Gamma^{0} \tau^{0} + i \sum_{a} \Gamma^{a} \tau^{a} \right]
\end{split}
\end{equation}
with $\Gamma^{0} = \sigma^{x}_{\textrm{pos}} \sigma^{0}_{m}$, $\Gamma^{a} =\sigma^{y}_{\textrm{pos}} \sigma^{a}_{m}$. The Hamiltonian for one plaquette can then be written down as
\begin{equation}
H_T=-J\text{Tr}\left[U(\vec{x},\hat{\mu})U(\vec{x}+\hat{\mu},\hat{\nu})U(\vec{x}+\hat{\mu}+\hat{\nu},-\hat{\mu}-\hat{\nu})\right].
\end{equation}
In order to identify all the terms of the trace one can use the condition $\textrm{tr}[\tau^\alpha\tau^\beta\tau^\gamma] \neq 0$. The condition is met if and only if
\begin{eqnarray}
\alpha = \beta = \gamma=0,\nonumber\\
\alpha \neq \beta \neq \gamma \neq 0,\nonumber\\
\alpha=0;\beta=\gamma\neq0\nonumber,\\
\beta=0;\alpha=\gamma\neq0\nonumber,\\
\gamma=0;\beta=\alpha\neq0.
\end{eqnarray}
The first condition gives $1$ term, the second $6$ terms, and $9$ terms for the last three equations, for a total of $16$ terms
\begin{equation}
\begin{split}
\textrm{Tr}\left[ U_{12} U_{23} U_{31} \right]=\frac{1}{4} \left\{ \Gamma^0_{12}\Gamma^0_{23}\Gamma^0_{31}+ \sum_{abc}\epsilon_{abc}\Gamma^a_{12}\Gamma^b_{23}\Gamma^c_{31} -\sum_{a}\Big[\Gamma^0_{12}\Gamma^a_{23}\Gamma^a_{31} + \Gamma^a_{12}\Gamma^0_{23}\Gamma^a_{31}+\Gamma^a_{12}\Gamma^a_{23}\Gamma^0_{31}\Big] \right\},
\end{split}
\end{equation}
where we have used the subscripts $12,23,31$ to represent the links $(\vec{x},\hat{\mu}),(\vec{x}+\hat{\mu},\hat{\nu}),(\vec{x}+\hat{\mu}+\hat{\nu},-\hat{\mu}-\hat{\nu})$, respectively.
The pure-gauge Hamiltonian can then be written in terms of $\Gamma$ matrices as
\begin{equation}
\label{ExplicitHam}
\begin{split}
H=-\frac{J}{2} \left\{ \Gamma^0_{12}\Gamma^0_{23}\Gamma^0_{31}+ \sum_{abc}\epsilon_{abc}\Gamma^a_{12}\Gamma^b_{23}\Gamma^c_{31} -\sum_{a}\Big[\Gamma^0_{12}\Gamma^a_{23}\Gamma^a_{31} + \Gamma^a_{12}\Gamma^0_{23}\Gamma^a_{31}+\Gamma^a_{12}\Gamma^a_{23}\Gamma^0_{31}\Big] \right\}.
\end{split}
\end{equation}
Then, one can explicitly write this Hamiltonian in terms of up to six-body interactions between qubits, for instance
\begin{equation}
\begin{split}
\Gamma^0_{12}\Gamma^0_{23}\Gamma^0_{31} =& \sigma^{x}_{\textrm{pos}12} \sigma^{x}_{\textrm{pos}23} \sigma^{x}_{\textrm{pos}31} \\
\Gamma^a_{12}\Gamma^b_{23}\Gamma^c_{31} =& \sigma^{y}_{\textrm{pos}12} \sigma^{y}_{\textrm{pos}23} \sigma^{y}_{\textrm{pos}31} \sigma^{a}_{m12} \sigma^{b}_{m23} \sigma^{c}_{m31} \\
\sum_{a}\Gamma^0_{12}\Gamma^a_{23}\Gamma^a_{31} =& \sigma^{x}_{\textrm{pos}12} \sigma^{y}_{\textrm{pos}23} \sigma^{y}_{\textrm{pos}31} \sum_{a} \sigma^{a}_{m23} \sigma^{a}_{m31}. 
\end{split}
\end{equation}

\subsection{Experimental setups}

In this section we provide a brief overview of the experimental setups considered for the implementation of the proposed simulator. The first setup considered in the main text is comprised of six transmon qubits provided with tunable coupling to a single microwave resonator. A transmon qubit is a special instance of a Cooper pair box~\cite{Bouchiat98S}, operating at specific regimes. 

A Cooper pair box is made of a piece of a superconducting island connected to a voltage source through a capacitor, and grounded via a Josephson junction. The temperature of these superconducting systems is low enough so that the quantum state of the setup can be defined by the number of Cooper pairs on the superconducting island. In order to add tunability and control to this quantum system, the single Josephson junction of the Cooper pair box is substituted with a SQUID loop, threaed with magnetic fluxes. These superconducting systems can be coupled to superconducting microwave resonators, which can be used to induce, via electromagnetic interactions, tunnelling of Cooper pairs in and out the superconducting island, as well as reading out the state of the system~\cite{Blais04S,Wallraff04S}. The Hamiltonian of the setup comprising a Cooper pair box coupled to a microwave resonator reads
\begin{eqnarray}
\label{CPB}
H_S=4E_{C}(n-n_{g})^2-E_{J}\cos\phi
+\omega_ra^{\dagger}a+2\beta eV_{\textrm{rms}}n(a+a^{\dagger}).\label{HT}
\end{eqnarray}
Here we have defined the quantized charge $n$ on the superconducting island, which is proportional to the number of Cooper pairs, the offset charge $n_g$ and the quantized flux $\phi$ that threads the SQUID loop. The $a$($a^{\dagger}$) are bosonic operators that act upon the resonator field, of frequency $\omega_r$. $E_{C}$ stands for the charging energy of the superconducting island, while $E_{J}=E_{J}^{\textrm{max}}|\cos(\pi\Phi_i/\Phi_0)|$ is the Josephson energy of the dc-SQUID loop. The coupling coefficient $\beta$ is a function of the circuit capacitances, $V_{\textrm{rms}}$ is the root mean square voltage of the resonator, and $e$ is the electron charge. 

One of the main problems that arise in using Cooper pair boxes as controlled quantum systems is the high sensitivity of their quantum state to noise on the offset charge $n_g$. By using large capacitive pads shunting the superconducting island, one can achieve insensitivity with respect to this noise. These Cooper pair boxes provided with large capacitive pads are known as transmon qubits~\cite{Koch07S}. In this parameter regime, the Hamiltonian of Eq.~(\ref{CPB}) models a transmon qubit. It can be rewritten in a diagonal basis for the part which is not coupled to the resonator,
\begin{align}
H_S\rightarrow\sum_{i=0}^2\omega_i \ket{i}\bra{i}+\omega_ra^{\dagger}a+\sum_{i=0}^2 g_{i,i+1}(\ket{i}\bra{i+1}+\ket{i+1}\bra{i})(a+a^{\dagger}),\label{HamcircuitQED}
\end{align}
where we have included the first three levels $i=\{0,1,2\}$ of this device. Higher levels are usually excluded from the quantum dynamics, as being more energetic and hard to access. The first three levels can be approximately modeled as the first three levels of an anharmonic oscillator, with a relative anharmonicity factor of $\alpha_r=(\omega_2-2\omega_1)/\omega_1$. If $\alpha_r$ is big enough, one can consider the transmon qubit device a two-level system~\cite{Koch07S}.

The setups envisioned in this work use two variants of the transmon design. Each of the six devices sketched in Fig.~1a in the main text is made of three superconducting islands, connected via two SQUID loops, marked with different colors in the figure. The magnetic fluxes acting on these loops modify the quantum level structure of the system, and an appropriate manipulation of these fluxes in time allows for a fast modulation of the effective coupling $g_{i,i+1}(t)$ to the resonator (see Eq.~\ref{HamcircuitQED}). This modulation does not affect the bare energy levels $\omega_i$ of the device~\cite{Gambetta11S,Srinivasan12S} and gives the possibility of performing red and blue sidebands upon the coupled qubit-resonator system, which, in turns, allows to generate many-body interactions~\cite{Mezzacapo14S}.

The setup considered in Fig.~1b in the main text describes transmon qubits with improved design, the so called Xmon qubits~\cite{Barends13S}. In this case, the shape of the capacitance pads is designed in order to improve coherence times of the device, by reducing coupling to material defects and radiative losses. The interactions among these qubits are mediated by direct capacitive couplings, not making use of a common superconducting resonator. For this reason, the many-body interactions are reduced in this case to nearest-neighbor two-qubit gates.

\subsection{Many-body interactions and gauge invariance for different sectors}

In order to simulate the Hamiltonian associated with the non-Abelian model in Eq.~(\ref{ExplicitHam}), one can simulate stepwise the single many-body monomials that compose the total interaction. Each of them has the form of an $N$-qubit string term
\begin{equation}
\label{GenericNbody}
H_N=\sigma^i_1\sigma^j_2\cdots\sigma^k_N,
\end{equation}
where the set of superindices can assume values $\{i,j,...k\}\in\{x,y,z\}$. Up to a maximum of $2N$ local rotations, this general interaction can be mapped back to, e.g., $\sigma^x_1\sigma^x_2\cdots\sigma^x_N$. In what follows, we will show how to obtain this effective interaction in two different scenarios. In one case, one can use a set of $N$ tunable-coupling transmon qubits coupled to a single microwave resonator. In the second scenario, a set of $N$ charge-like qubits, e.g., transmon or Xmon qubits, are coupled to an additional ancillary qubit, for a total of $N+1$ qubits. 

We consider the sequence of collective and single qubit gates~\cite{Mueller11S,Mezzacapo14S}, as 
\begin{equation}
\label{UMS}
U_\textrm{S}(t)=e^{i(\pi/4)\sum_{i<j}\sigma^x_i\sigma^x_j}e^{igt\sigma_1^z}e^{-i(\pi/4)\sum_{i<j}\sigma^x_i\sigma^x_j},
\end{equation}
and obtain the set of interactions 
\begin{align}
U_\textrm{S}(t)=&e^{-igt\sigma^z_1\sigma_{2}^{x}\cdots\sigma_{N}^{x}}, N=4n-1,\nonumber\\
U_\textrm{S}(t)=&e^{igt\sigma^z_1\sigma_{2}^{x}\cdots\sigma_{N}^{x}}, N=4n+1,\nonumber\\
U_\textrm{S}(t)=&e^{-igt\sigma^y_1\sigma_{2}^{x}\cdots\sigma_{N}^{x}}, N=4n,\nonumber\\
U_\textrm{S}(t)=&e^{igt\sigma^y_1\sigma_{2}^{x}\cdots\sigma_{N}^{x}}, N=4n-2.\label{stabilizer}
\end{align}
The collective interaction $e^{i(\pi/4)\sum_{j<i}\sigma^x_i\sigma^x_j}$ in Eq.~(\ref{UMS}) can be realized within a set of $N$ three-island transmon qubits, by threading the two SQUID loops embedded in each device with fast-tunable magnetic fluxes, tuned at appropriate frequencies~\cite{Mezzacapo14S}. The general $N$-body term of Eq.~(\ref{GenericNbody}) can be simulated with this approach with a total of $2$ collective gates, and a number of single qubit gates which is upper bounded by $2N+1$. The Hamiltonian in Eq.~(\ref{ExplicitHam}) is composed of $16$ monomials, $1$ three-body term, $6$ six-body terms, and $9$ five-body terms. Therefore, the simulation for one digital step will amount to implement $32$ collective gates and a number of single-qubit gates which is bounded by $184$.

\begin{figure}
\includegraphics[scale=0.55]{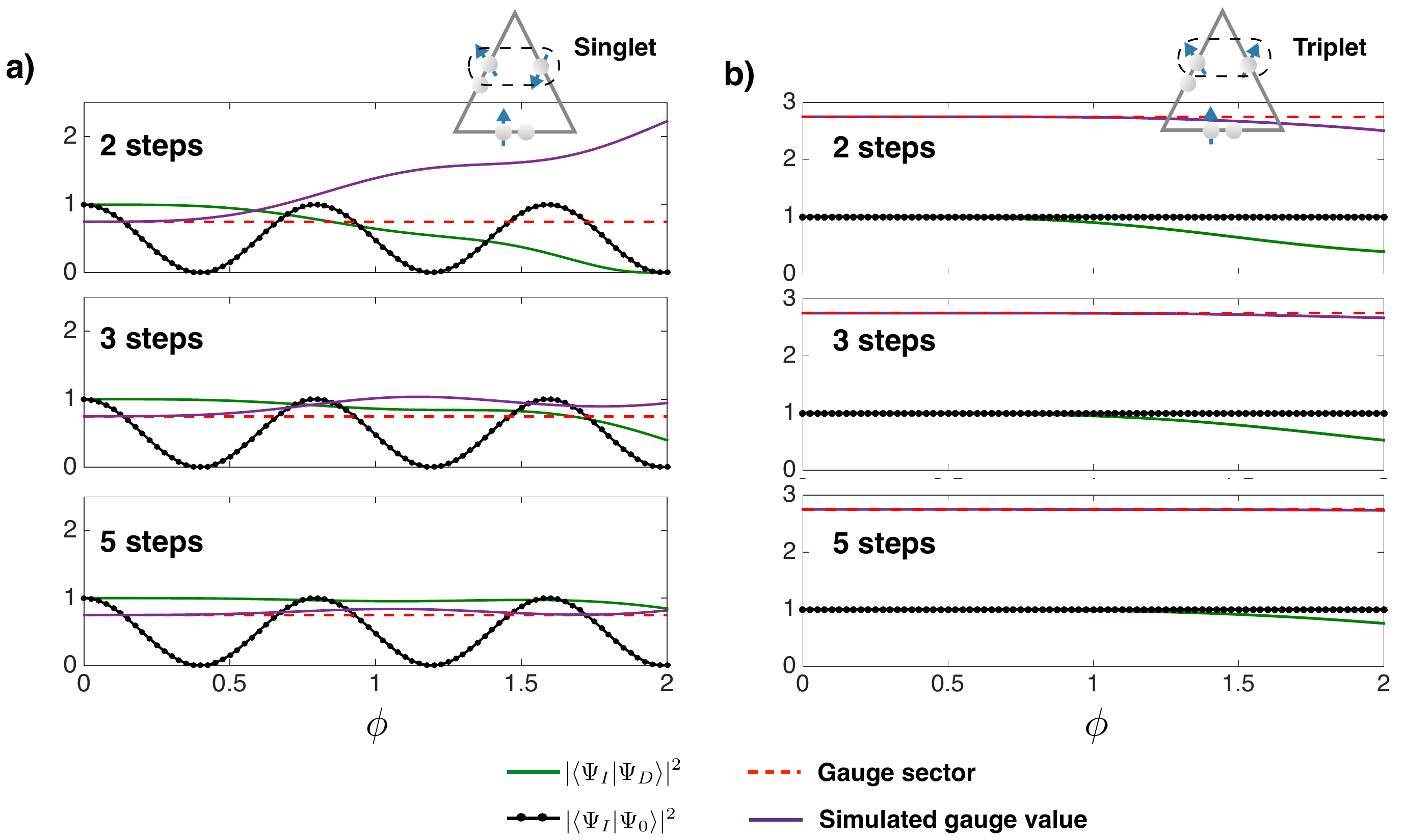}
\caption{(Color online) Gauge invariance is preserved through digitization. The simulated gauge value  $\langle\sum_{\vec{x}} {\vec G^{2}({\vec x })}\rangle_D$ converges to the expected invariant, $3/4$ for a) and $11/4$ for b), as the number of digital steps increases. The initial states $\ket{\Psi_0}$  chosen in a) and b) are represented in the main text, inFig. 2 a) and c), respectively. The overlap $|\langle\Psi_I|\Psi_0\rangle|^2$ witnesses the periodic oscillations of the initial state on the triangular plaquette for the first case, while it remains invariant in the second one.\label{SinTrip}}
\end{figure}

One can implement the interaction $H_N$ in a set of capacitively-coupled charge-like qubits, by decomposing it into a set of $C$-phase and single-qubit gates. First, let us consider the sequence of gates in Eq.~(\ref{UMS}) between $N$ qubits and an additional ancillary one, labeled by $A$. We consider also that the gates are rotated about the $Y$ axis of an angle $\pi/4$. The sequence of gates in Eq.~(\ref{UMS}) can then be rewritten in the following way
\begin{equation}
\label{UMSR}
U_\textrm{S,A}(t)=e^{i(\pi/4)\sigma^z_A\sum_{j=1}^N\sigma^z_j}e^{-igt\sigma_A^x}e^{-i(\pi/4)\sigma^z_A\sum_{j=1}^N\sigma^z_j}.
\end{equation}
This sequence of gates is equivalent to the following interactions 
\begin{align}
U_\textrm{S,A}(t)=&e^{ igt\sigma^x_A\sigma_{1}^{z}\cdots\sigma_{N}^{z}}, N=4n-1,\nonumber\\
U_\textrm{S,A}(t)=&e^{-igt\sigma^x_A\sigma_{1}^{z}\cdots\sigma_{N}^{z}}, N=4n+1,\nonumber\\
U_\textrm{S,A}(t)=&e^{-igt\sigma^y_A\sigma_{1}^{z}\cdots\sigma_{N}^{z}}, N=4n,\nonumber\\
U_\textrm{S,A}(t)=&e^{igt\sigma^y_A\sigma_{1}^{z}\cdots\sigma_{N}^{z}}, N=4n-2.\label{stabilizer}
\end{align}
If the ancillary qubit is initialized in an eigenstate of $\sigma^x_A$($\sigma^y_A$), $\ket{\Psi_A}=(\ket{\uparrow}_A+\ket{\downarrow}_A)/\sqrt{2}$ ($\ket{\Psi_A}=(\ket{\uparrow}_A-i\ket{\downarrow}_A)\sqrt{2}$), then the effective dynamics on the $N$-qubit system would be equivalent to the interaction $e^{-igt\sigma_{1}^{z}\cdots\sigma_{N}^{z}}$.
The interaction $e^{i(\pi/4)\sigma^z_A\sum_{j=1}^N\sigma^z_j}$ can be realized as a sequence of $N$ $ZZ$ gates between the ancillary qubit and all the other ones, $e^{i(\pi/4)\sigma^z_A\sum_{j=1}^N\sigma^z_j}=\prod_{i=1}^{N}e^{i(\pi/4)\sigma^z_A\sigma_i^z}$. Each of these two-body gates can be realized as a sequence of two local $Z$-rotations and a $C$-phase gate between the qubits $1$ and $i$, $CP_{1i}(\phi)$. In fact, defining
\begin{eqnarray}
e^{i\phi/2}e^{-i(\phi/2)\sigma^z}=\left(\begin{array}{cc} 1 & 0 \\0 & e^{i\phi}\end{array}\right)\!\!, \,\,CP_{1i}(\phi)=\left(\begin{array}{cccc} 1 & 0 & 0& 0\\0 & 1 & 0 & 0\\ 0 & 0 & 1 & 0\\ 0 & 0 & 0 & e^{-i2\phi}\end{array}\right)\!\!\label{EqCZ},
\end{eqnarray}
one has that 
\begin{equation}
e^{-i(\phi/2)\sigma^z_1}e^{-i(\phi/2)\sigma^z_i}CP_{1i}(\phi)=e^{i(\phi/2)\sigma_1^z\sigma_i^z}.
\end{equation}
In this way, an arbitrary $N$-body term can be encoded, up to a number of local rotations bounded by $2N$, with a sequence of $2N$ $C$-Phase gates and a number of single-qubit gates bounded by $6N+1$. Thus the quantum simulation of one digital step of the Hamiltonian in Eq.~(\ref{ExplicitHam}) will amount to realize $168$ $C$-Phase gates and a number of single-qubit rotations which is bounded by $520$.

The $C$-Phase gate between two qubits can be performed by adiabatically changing the frequency of one qubit, bringing the level $\ket{11}$ close to the anharmonic excitated level $\ket{02}$, to induce state-dependent phase accumulations. Unwanted errors due to leakage to the $\ket{02}$ state are exponentially suppressed in the gate duration. For finite gate times errors can be minimized upon optimization in the Fourier spectrum of the control signal~\cite{Martinis14S}. This optimization has been carried out for $CZ$ gates in~\cite{Barends14S}. For a generic quantum simulation involving $C$-Phase gates, one should define the set of angles to be used in the experiment, and optimize over the duration and Fourier spectrum of the gate, to avoid unwanted leakage. This has been done in~\cite{Barends15S} by increasing the gate time from $40$ ns to $55$ ns, not observing significant state leakages up to phases of $4$ rads. Additional errors on the fidelity of the gate come from standard decoherence of the qubits, and control errors, see Ref.~\cite{Barends14S} for a discussion of the error budget. 

\begin{figure}
\includegraphics[scale=0.3]{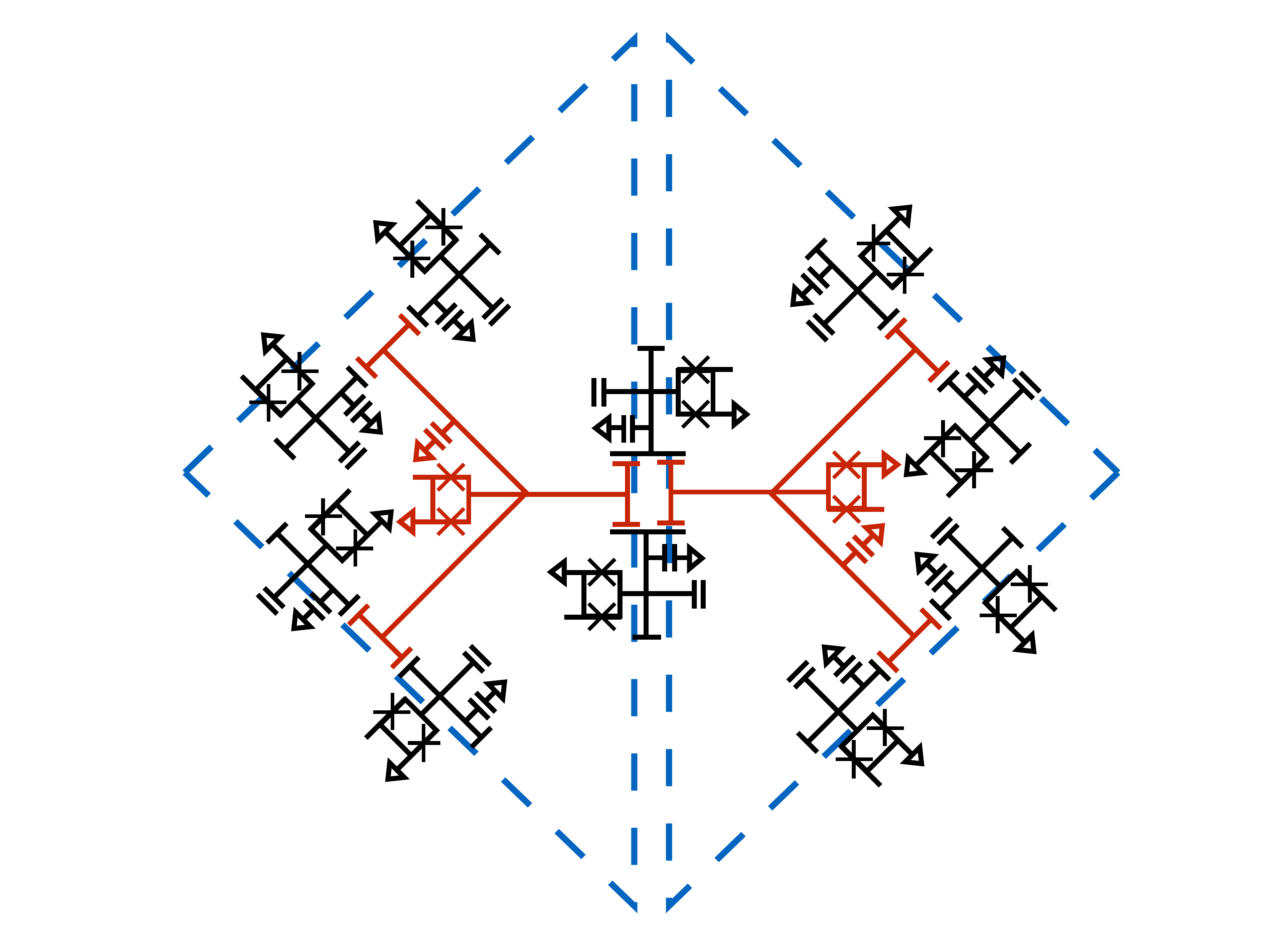}
\caption{(Color online) Scheme for an experiment involving two pure-gauge plaquettes, implemented in a lattice of capacitively coupled Xmon qubits. Multiple capacitive connections in superconducting pads with appropriate design can provide connectivity for larger lattices.\label{LargeLattice}}
\end{figure}

Finally, to complete the information given in the main text, we show in this section, in Fig.~\ref{SinTrip}, additional plots that complement Fig.~3 in the main text. The simulated gauge invariance is shown to converge for the singlet and triplet configurations on a single triangular plaquette. 

The proposed experiment can be extended to larger simulated lattices, which are implementable by considering setups of superconducting qubits with enhanced connectivity. The required connectivity can be achieved by using several microwave resonators, acting as non-local quantum buses, or by considering superconducting pads that can provide with multiple capacitive couplings, as schematized in Fig.~\ref{LargeLattice}. For instance, we can envision, as shown in the main text, a ladder architecture that can support a singlet gauge invariant subspace, where non-Abelian string dynamics could be studied.
\pagebreak

\subsection{Scaling of errors for larger lattices}

It has been shown~\cite{Berry07S} that the scaling of the number of steps $N_s$ needed to achieve a digital error $\epsilon$ in a Suzuki-Lie-Trotter decomposition is almost linear in the simulated phase of the dynamics and sub-polinomial in $\epsilon$, according to the formula 
\begin{equation}
N_{s}\leq \frac{2m5^{2k}(m||H||t)^{1+1/2k}}{\epsilon^{1/2k}}.
\end{equation}  
In the formula $m$ is the number of terms that compose a generic Hamiltonian $H=\sum_{i=1}^m h_i$, and $k$ is the fractal degree of approximation~\cite{Suzuki90S}, which we set to $k=1$. For one plaquette one has $m=16$, $1$ three-body term, $6$ six-body terms, and $9$ five-body terms, all present in the system with the energy $J/2$. Considering that the norm of an arbitrary product of $P$ Pauli matrices is $||\sigma_1^j\sigma_2^i\cdots\sigma_P^l||=1$, $\{i,j,l\}\in \{x,y,z\}$, one has, using a triangular inequality, that the norm of the $SU(2)$ Hamiltonian for one plaquette is bounded by $||H||\leq|J|/8$. As a consequence, the number of steps necessary to implement a dynamics with error $\epsilon$ for the time $t$ on a lattice of $N$ plaquettes is bounded by
\begin{equation}
\label{ErrBound}
N_{s}\leq\frac{2300N(N|J|t)^{3/2}}{\sqrt{\epsilon}}.
\end{equation}
Notice that this formula gives a very loose bound which is not dependent on the underlying structure of the Hamiltonian. Nevertheless, it shows that the scaling of the number of steps is efficiently solvable by a quantum simulator, not being exponential in the resources used. The exact scaling of the number of steps cannot be known (up to an exact solution of the problem, which is not addressable numerically for larger lattices), however it is useful to extract the dependence $N_s\sim N^{5/2}$ from the upper-bound formula in Eq.~(\ref{ErrBound}).
Assuming independent experimental errors on the single Trotter steps, one can relate the total number of steps $N_s$ to the required experimental fidelity for the implementation of a single Hamiltonian term. We impose an accumulated experimental error $N_s\epsilon_s$ of the same order of magnitude of the Trotter error $N_s \epsilon_s\sim\epsilon$. This enforces the error on the single step to be of the order $\epsilon_s \sim \epsilon/N_s$.
Besides these issues on error scaling and accumulation, the accesible simulation time has to be smaller than $T_1$ and $T_2$ of the single qubits, defining a system of necessary conditions for the simulation of a lattice of $N$ plaquettes
\begin{equation}
\begin{cases}
N_{s}\leq\frac{2300N(N|J|t)^{3/2}}{\sqrt{\epsilon}}\\
\epsilon_s\sim \frac{\epsilon}{  N_s}\\
t\ll T_1,T_2.
\end{cases}
\end{equation}

\subsection{Matter-field $SU(2)$ Gauge Interactions}

In the main text we have focused on pure gauge dynamics with plaquette interactions, where usual perturbative approaches do not get energy scales which are big enough. In this section, we point out the basic elements needed for the simulation of the simplest $SU(2)$ gauge model coupled to a matter field in $(1+1)$-dimensions. The basic ingredients and the validity of the different energy-scale regimes have been already discussed in the simulation of Abelian gauge models in trapped ions and superconducting circuits \cite{haukeS,rablS}. This section can be seen as a generalization to non-Abelian setups.

The ideal interaction Hamiltonian that we would like to simulate is given by
\begin{equation}
\begin{split}
H_{\text{int}} &= \sum_{i} b^{\dagger}(i) U(i,i+1) b(i+1) + \text{H.c.} \\
&= \sum_{i \alpha \beta} b^{\dagger}_{\alpha}(i) \left\{ \left[ c_{\alpha L}(i) c^{\dagger}_{\beta R}(i+1) \right] + \sum_{\mu \nu}  \left[ \sigma^{y}_{\alpha \mu} c^{\dagger}_{\mu L}(i) c_{\nu R} (i+1) \sigma^{y}_{\nu \beta} \right] \right\} b_{\beta}(i+1) + \text{H.c.},
\end{split}
\end{equation} 
where $b_{\alpha}(i)$ and $c_{\alpha p}(i)$ are hard-core bosonic modes, with $\alpha \in \{ \uparrow, \downarrow \}$, $p \in \{ L, R \}$ and $i \in \mathbb{Z}$.

In terms of these modes, the former Hamiltonian has two local (gauge) symmetries:
\begin{itemize}
\item $SU(2)$ gauge invariance around every site with generators: $G^{a}(i) =R^{a}(i-1,i) + \frac{1}{2} \sum_{\alpha \beta} b^{\dagger}_{\alpha}(i) \sigma^{a}_{\alpha \beta} b_{\beta}(i) + L^{a}(i,i+1)  $
\item $U(1)$ gauge invariance on the links with generator: $G(i,i+1) = \sum_{\alpha} \left[ c^{\dagger}_{\alpha L}(i) c_{\alpha L}(i) + c^{\dagger}_{\alpha R}(i+1) c_{\alpha R}(i+1) \right] $
\end{itemize}

One way to obtain the previous Hamiltonian is to start with operators that respect one of the symmetries at the fundamental level and introduce an energy penalty for the combinations of operators that do not fulfill the second symmetry. In general, it is simpler to enforce the Abelian $U(1)$ symmetry, since it only requires density-density interaction, and start with interactions that fulfill the non-Abelian symmetry from the beginning.

The basic building blocks are given by a strong interaction that enforces the Abelian symmetry, i.e.,
\begin{equation}
\label{H00}
H_{0} = \omega \sum_{i \alpha}  \left[ b^{\dagger}_{\alpha}(i) b_{\alpha}(i) + \sum_{p} c^{\dagger}_{\alpha p}(i) c_{\alpha p}(i) \right] + \Omega \sum_{i} \left[ G(i,i+1)  - N_{0} \right]^{2}.
\end{equation}
Here, the $\omega$ frequency of the bosonic modes splits the Hilbert space into subspaces with an equal number of excitations. Performing a rotating-wave approximation,  interactions that conserve the number of particles are energetically allowed, while the term with energy scale $\Omega$ enforces $N_{0}$ excitations per link.

On top of the previous Hamiltonian of Eq.~(\ref{H00}), a perturbative $SU(2)$ invariant hopping term is applied
\begin{equation}
V=J_0 \sum_{i \alpha} \left\{ b^{\dagger}_{\alpha}(i) c_{\alpha L}(i) + c^{\dagger}_{\alpha R}(i) b_{\alpha}(i) +  \sum_{\mu} \left[ b^{\dagger}_{\alpha}(i) \sigma^{y}_{\alpha \mu} c^{\dagger}_{\mu L}(i) + c_{\mu R} (i) \sigma^{y}_{\mu \alpha} b_{\alpha}(i)  \right] \right\} + \text{H.c.}
\end{equation}

Hence, in second-order perturbation theory, up to a renormalization of the bare frequencies $\omega$ of the modes, the effective Hamiltonian is given by
\begin{equation}
\begin{split}
H_{\text{eff}} =&-2 \frac{J^{2}_0}{\Omega} \sum_{i \alpha \beta} b^{\dagger}_{\alpha}(i) \left\{ \left[ c_{\alpha L}(i) c^{\dagger}_{\beta R}(i+1) \right] + \sum_{\mu \nu}  \left[ \sigma^{y}_{\alpha \mu} c^{\dagger}_{\mu L}(i) c_{\nu R} (i+1) \sigma^{y}_{\nu \beta} \right] \right\} b_{\beta}(i+1) + \text{H.c.} \\
&-2 \frac{J^{2}_0}{\Omega} \sum_{i} \left\{ \sum_{\alpha} b^{\dagger}_{\alpha}(i) b_{\alpha}(i)  \sum_{\beta} \left[ c^{\dagger}_{\beta R}(i) c_{\beta R}(i) + c^{\dagger}_{\beta L}(i) c_{\beta L}(i) \right] \right\},
\end{split}
\end{equation}
which reduces to the ideal Hamiltonian $H_{\text{int}}$, up to a density-density term that does not break any of the symmetries and it could be reduced or even eliminated, allowing a counter-term of the same type in the initial Hamiltonian $H_{0}$.

\end{widetext}

\end{document}